\documentclass{article}

\PassOptionsToPackage{numbers, compress}{natbib}

\usepackage[preprint]{neurips_2023}

\usepackage[utf8]{inputenc} 
\usepackage[T1]{fontenc}    
\usepackage{hyperref}       
\usepackage{url}            
\usepackage{booktabs}       
\usepackage{amsfonts}       
\usepackage{nicefrac}       
\usepackage{microtype}      
\usepackage{xcolor}         

\usepackage{amsmath,amsfonts,bm}

\def\eqref#1{equation~\ref{#1}}

\def\1{\bm{1}}

\def\vf{{\bm{f}}}

\def\vm{{\bm{m}}}

\def\vu{{\bm{u}}}
\def\vv{{\bm{v}}}
\def\vw{{\bm{w}}}
\def\vx{{\bm{x}}}
\def\vy{{\bm{y}}}

\def\mC{{\bm{C}}}

\def\mI{{\bm{I}}}

\def\mM{{\bm{M}}}
\def\mN{{\bm{N}}}

\def\mU{{\bm{U}}}

\def\mX{{\bm{X}}}
\def\mY{{\bm{Y}}}

\def\mSigma{{\bm{\Sigma}}}

\DeclareMathAlphabet{\mathsfit}{\encodingdefault}{\sfdefault}{m}{sl}
\SetMathAlphabet{\mathsfit}{bold}{\encodingdefault}{\sfdefault}{bx}{n}

\DeclareMathOperator*{\argmax}{arg\,max}

\usepackage{hyperref}
\usepackage{url}
\usepackage{graphicx}
\graphicspath{ {./figures/} }
\usepackage{amsmath}
\usepackage{cleveref}

\usepackage{algorithm}
\usepackage{algpseudocode}

\title{Factorized Discriminant Analysis for Genetic Signatures of Neuronal Phenotypes}

\author{%
Mu Qiao\textsuperscript{1,2}\\
\textsuperscript{1}Division of Biology and Biological Engineering, \\
California Institute of Technology, Pasadena, CA 91125\\
\textsuperscript{2}Present address: LinkedIn, Mountain View, CA, 94043\\
\texttt{muqiao0626@gmail.com}
}

\begin{document}

\maketitle

\begin{abstract}
Navigating the complex landscape of single-cell transcriptomic data presents significant challenges. Central to this challenge is the identification of a meaningful representation of high-dimensional gene expression patterns that sheds light on the structural and functional properties of cell types. Pursuing model interpretability and computational simplicity, we often look for a linear transformation of the original data that aligns with key phenotypic features of cells. In response to this need, we introduce factorized linear discriminant analysis (FLDA), a novel method for linear dimensionality reduction. The crux of FLDA lies in identifying a linear function of gene expression levels that is highly correlated with one phenotypic feature while minimizing the influence of others. To augment this method, we integrate it with a sparsity-based regularization algorithm. This integration is crucial as it selects a subset of genes pivotal to a specific phenotypic feature or a combination thereof. To illustrate the effectiveness of FLDA, we apply it to transcriptomic datasets from neurons in the Drosophila optic lobe. We demonstrate that FLDA not only captures the inherent structural patterns aligned with phenotypic features but also uncovers key genes associated with each phenotype.
\end{abstract}

\section{Introduction}
\label{introduction}

The analysis of gene expression data in single cells presents an intriguing and complex problem. Each cell's gene expression data can be viewed as a high-dimensional vector, allowing each cell to be represented as a single point in the vast space of gene expression. Clusters form within this space, each identifiable and associated with a particular cell type, thanks to the verification from the molecular markers of cell types \citep{macoskoHighlyParallelGenomewide2015, tasicAdultMouseCortical2016, shekharCOMPREHENSIVECLASSIFICATIONRETINAL2016, tasicSharedDistinctTranscriptomic2018, pengMolecularClassificationComparative2019}.

When the phenotypic traits of each cell type are known, either from past studies or direct measurement \citep{sanesTypesRetinalGanglion2015, zengNeuronalCelltypeClassification2017, cadwellElectrophysiologicalTranscriptomicMorphologic2016, strellPlacingRNAContext2019}, we can label each cell type according to its unique characteristics. For example, differentiation of neuronal cell types could be achieved through analyzing a variety of features, such as dendritic and axonal laminations, electrophysiological properties, and connectivity \citep{sanesTypesRetinalGanglion2015, zengNeuronalCelltypeClassification2017, gouwensClassificationElectrophysiologicalMorphological2019}. These features are often categorical in nature.

A critical challenge arises when we attempt to factorize the high-dimensional gene expression data into modules that align with these phenotypes. In simple terms, we aim to find a low-dimensional embedding of gene expression where each axis signifies a single factor. This factor might correspond to a specific phenotypic feature or potentially, the combination of several.

Ideally, variation along one axis in the embedding space would exclusively affect one phenotypic feature. However, due to inevitable noise in the data, this is challenging to achieve. As a workaround, we allow for data projected along one axis to vary primarily with one phenotypic feature and minimally with others. Simultaneously, we want to preserve cell type identities in the low-dimensional space. This means that cells of the same type should remain in close proximity within the embedding space, while cells of different types remain distinct.

In order to address this issue, we propose the method of factorized linear discriminant analysis (FLDA). This is a supervised dimensionality reduction technique, rooted in the concepts of multi-way analysis of variance (ANOVA) \citep{fisherCorrelationRelativesSupposition1918}. FLDA enables the factorization of data into components that correspond to phenotypic features and their combinations. It then seeks a linear transformation that is highly variable with one component, yet stable with others. The power of this approach lies in its simplicity and interpretability. To further leverage our analysis, we introduce a sparse variant of this method. This variant restricts the number of non-zero elements contributing to each linear projection, thereby identifying a subset of genes crucial to each phenotype. The efficacy of FLDA is demonstrated through its application to a single-cell RNA-Seq dataset of T4/T5 neurons in Drosophila \citep{kurmangaliyevModularTranscriptionalPrograms2019}, focusing particularly on two phenotypes: dendritic location and axonal lamination.

\section{Factorized Linear Discriminant Analysis (FLDA)}
\label{FLDA}

Let's consider a situation where each cell type can be characterized by two phenotypic features, both of which are categorical. This essentially means that the sample space for cell types is a Cartesian product of the sample spaces of the two phenotypic features $I$ and $J$:

\begin{equation} \label{eq0}
I \times J = \{(i, j)|i \in I, j \in J\}
\end{equation}

In this equation, $i$, $j$ represent different categories of the two phenotypic features. Suppose we have observed $n_{ij}$ cells for each cell type $(i,j)$. This information can be visualized with a contingency table, as shown in Figures 1A and 1B. Note here we account for the scenario where the table might be only partially filled.

\begin{figure}
\centering
\includegraphics[trim=0 525 0 100, clip, width=0.95\linewidth]{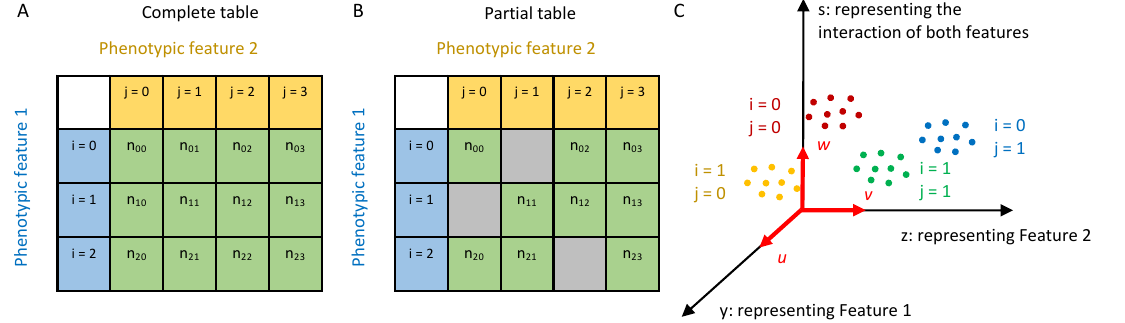}
\caption{Illustration of our approach. (A,B) Here, cell types are represented by two phenotypic features, labeled with $i$ and $j$ respectively. If only some combinations of the two features are observed, we have a partial contingency table (B) rather than a complete one (A). (C) We aim to find linear projections of the data that separate the cell types in a manner factorized according to the two features. In this diagram, $\vu$, $\vv$, and $\vw$ are aligned with Feature 1, Feature 2, and their combination respectively, with the projected coordinates $y$, $z$, and $s$.}
\label{fig1}
\end{figure}

We denote the expression values of $g$ genes measured in the $k$th cell of the cell type $(i,j)$ as $\vx_{ijk} (k \in {1,2,...n_{ij}})$ ($\vx_{ijk} \in \mathbf{R}^{g})$. Our task is to find linear projections $y_{ijk} = \vu^T \vx_{ijk}$ ($\vu \in \mathbf{R}^{g})$ and $z_{ijk} = \vv^T \vx_{ijk}$ ($\vv \in \mathbf{R}^{g})$ that align with features $i$ and $j$ respectively (see Figure 1C).

To address this, we explored whether we could factorize, for example, $y_{ijk}$, into components dependent on features $i$ and $j$. By employing the principles of linear factor models from multi-way ANOVA and the concept of variance partitioning, we formulated an objective function to find $\vu^{}$ that maximizes this objective (for a detailed analysis, refer to Appendix A).

\begin{equation} \label{eq1}
\vu^{*} = \argmax_{\vu\in \mathbf{R}^{g}} \frac{\vu^T \mN_{A} \vu}{\vu^T \mM_{e} \vu}
\end{equation}

With a complete table, where $a$ and $b$ are the number of categories for feature $i$ and $j$, we have:

\begin{equation} \label{eq2}
\mN_{A} = \mM_{A} - \lambda_{1}\mM_{B} - \lambda_{2}\mM_{AB}
\end{equation}

Here, $\mM_{A}$, $\mM_{B}$, and $\mM_{AB}$ denote the covariance matrices explained by feature $i$, feature $j$, and their combination respectively. The hyper-parameters $\lambda_{1}$ and $\lambda_{2}$ determine the relative weights of $\mM_{B}$ and $\mM_{AB}$ in comparison to $\mM_{A}$. The residual covariance matrix, $\mM_{e}$, represents variance within cell type clusters and signifies noise in gene expressions. The formal definitions of these terms are as follows:

\begin{equation} \label{eq3}
\mM_{A} = \frac{1}{a-1}\sum_{i=1}^{a}(\vm_{i.} - \vm_{..})(\vm_{i.} - \vm_{..})^T
\end{equation}

\begin{equation} \label{eq4}
\mM_{B} = \frac{1}{b-1}\sum_{j=1}^{b} (\vm_{.j} - \vm_{..})(\vm_{.j} - \vm_{..})^T
\end{equation}

\begin{equation} \label{eq5}
\begin{split}
\mM_{AB} = \frac{1}{(a-1)(b-1)}\sum_{i=1}^{a}\sum_{j=1}^{b} (\vm_{ij} - \vm_{i.} - \vm_{.j} + \vm_{..})\
(\vm_{ij} - \vm_{i.} - \vm_{.j} + \vm_{..})^T
\end{split}
\end{equation}

\begin{equation} \label{eq6}
\mM_{e} = \frac{1}{N-ab}\sum_{i=1}^{a}\sum_{j=1}^{b}[\frac{1}{n_{ij}}\sum_{k=1}^{n_{ij}} (\vx_{ijk} - \vm_{ij})(\vx_{ijk} - \vm_{ij})^T]
\end{equation}

and

\begin{equation} \label{eq7}
\vm_{..} = \frac{1}{ab}\sum_{i=1}^{a} \sum_{j=1}^{b} \vm_{ij}
\end{equation}

\begin{equation} \label{eq8}
\vm_{i.} = \frac{1}{b}\sum_{j=1}^{b} \vm_{ij}
\end{equation}

\begin{equation} \label{eq9}
\vm_{.j} = \frac{1}{a}\sum_{i=1}^{a} \vm_{ij}
\end{equation}

with

\begin{equation} \label{eq10}
\vm_{ij} = \frac{1}{n_{ij}}\sum_{k=1}^{n_{ij}} \vx_{ijk}
\end{equation}

Analogously, the linear projections $\vv^{}$ for feature $j$ and $\vw^{}$ for the combination of both features $i$ and $j$ can be determined by similar formulas. By applying the same rationale to a partial table, we can derive $\vu^{}$ or $\vv^{}$ as the linear projection for feature $i$ or $j$ (see Appendix B for a detailed mathematical discussion).

Note that $\mN_{A}$ is symmetric and $\mM_{e}$ is positive definite, transforming the optimization problem into a generalized eigenvalue problem \citep{ghojoghEigenvalueGeneralizedEigenvalue2019}. When $\mM_{e}$ is invertible, $\vu^{*}$ is the eigenvector associated with the highest eigenvalue of $\mM_{e}^{-1} \mN_{A}$. Generally, if we aim to embed $\vx_{ijk}$ into a $d$-dimensional subspace aligned with feature $i$ ($d<a$), we take the eigenvectors corresponding to the $d$ largest eigenvalues of $\mM_{e}^{-1} \mN_{A}$, which we term as the top $d$ factorized linear discriminant components (FLDs).

In situations where the number of genes greatly exceeds the number of cells, $\mM_{e}$ becomes singular and non-invertible. In such cases, we resort to solutions suggested in \citep{friedmanRegularizedDiscriminantAnalysis1989, dudoitComparisonDiscriminationMethods2002, bickelTheoryFisherLinear2004} that uses a diagonal estimate of $\mM_{e}$: $diag(\hat\sigma_{1}^{2}, \hat\sigma_{2}^{2}, ..., \hat\sigma_{p}^2)$, where $\hat\sigma_{i}^{2}$ is the $i$th diagonal element of $\mM_{e}$. This solution has been employed in multiple computational biology studies \citep{tibshiraniClassPredictionNearest2003, butlerIntegratingSinglecellTranscriptomic2018,stuartComprehensiveIntegrationSingleCell2019}.

As multi-way ANOVA can handle contingency tables with more than two dimensions, our analysis can be easily extended to handle more than two phenotypic features \citep{hahn2023evolution}. In summary, FLDA is well-suitable for data whose labels form a Cartesian product of multiple features.

\section{Sparse Regularization of FLDA}
\label{SFLDA}

In computational biology applications, we are often interested in identifying a small subset of genes that effectively characterizes a specific phenotypic feature. This leads to the identification of axes with a few non-zero elements. To find such a sparse solution, we address the following optimization problem:

\begin{equation} \label{eq11}
\begin{aligned}
\vu^{*} = \argmax_{\vu\in \mathbf{R}^{g}} \frac{\vu^T \mN_{A} \vu}{\vu^T \mM_{e} \vu}
& & \text{subject to}
& & || \vu ||_0 \leq l
\end{aligned}
\end{equation}

where the number of non-zero elements of $\vu^{*}$ is constrained to be less or equal to $l$.

This problem, also known as a sparse generalized eigenvalue problem, presents three challenges \citep{tanSparseGeneralizedEigenvalue2018}: Handling extremely high-dimensional data, $\mM_{e}$ can be singular and non-invertible; Working with the normalization term $\vu^T \mM_{e} \vu$, which restricts the application of many sparse eigenvalue solutions; Maximizing a convex objective over a nonconvex set, a problem known to be NP-hard.

To overcome these challenges, we employ the truncated Rayleigh flow (Rifle) method, which was designed specifically for solving sparse generalized eigenvalue problems. The Rifle algorithm is a two-step process \citep{tanSparseGeneralizedEigenvalue2018}: First, it acquires an initial vector $\vu_{0}$ that is close to $\vu^{*}$. For this, we use the non-sparse FLDA solution as an initial estimate for $\vu_{0}$; Second, it iteratively performs a gradient ascent on the objective function. This is followed by a truncation step that retains the $l$ entries of $\vu$ with the highest values and sets the remaining entries to zero. The step-by-step process of applying the Rifle method to solve our problem is detailed in the following pseudo-code:

\begin{algorithm}
    \begin{algorithmic}
        \Procedure{Rifle}{$\mN_{A}, \mM_{e}, \vu_{0}, l, \eta$} \Comment{$\eta$ is the step size}
            \State $t = 1$
            \Comment{$t$ indicates the iteration number}
            \While{not converge}
            \Comment{Converge when $\vu_{t} \simeq \vu_{t-1}$}
                \State $\rho_{t-1} \gets \frac{\vu_{t-1}^T \mN_{A} \vu_{t-1}}{\vu_{t-1}^T \mM_{e} \vu_{t-1}}$
                \State $\mC \gets \mI + (\frac{\eta}{\rho_{t-1}})(\mN_{A} - \rho_{t-1}\mM_{e})$
                \State $\vu_{t} \gets \frac{\mC\vu_{t-1}}{|| \mC\vu_{t-1} ||_2}$
                \State Truncate $\vu_{t}$ by preserving the top $l$ entries of $\vu$ with the largest values and setting the remaining entries to 0
                \State $\vu_{t} \gets \frac{\vu_{t}}{|| \vu_{t} ||_2}$
                \State $t \gets t+1$
            \EndWhile
            \State \textbf{return} $\vu_{t}$
        \EndProcedure
    \end{algorithmic}
\end{algorithm}

As previously demonstrated in \citep{tanSparseGeneralizedEigenvalue2018}, the Rifle method can effectively converge to the unique sparse leading generalized eigenvector, assuming it exists, at the optimal statistical rate of convergence. The computational complexity of the second step in each iteration is $O(lg+g)$, indicating that the Rifle algorithm scales linearly with $g$, the number of genes in the input data.

In terms of hyperparameter selection, the step size $\eta$ should be small enough to ensure convergence, specifically $\eta \lambda_{max}(\mM_{e}) < 1$, where $\lambda_{max}(\mM_{e})$ is the largest eigenvalue of $\mM_{e}$. This is akin to taking small steps to ensure that we don't overshoot the optimal solution. The other hyperparameter, $l$, which determines the number of genes to be preserved, is chosen empirically based on the design of the subsequent experiment. This parameter can be adjusted depending on the specific requirement of a biological study.

\section{Related Work: Dimensionality Reduction}
\label{relatedMethod}

FLDA is one method for linear dimensionality reduction \citep{cunninghamLinearDimensionalityReduction2015}. In formal terms, linear dimensionality reduction can be defined as follows: Given $n$ data points, each of $g$ dimensions, $\mX = [\vx_{1}, \vx_{2}, ..., \vx_{n}] \in \mathbf{R}^{g \times n}$, and a chosen reduced dimensionality $r<g$, an objective function $f(.)$ is optimized to produce a linear projection $\mU \in \mathbf{R}^{r \times g}$. The result is a low-dimensional transformed dataset $\mY = \mU\mX \in \mathbf{R}^{r \times n}$.

Leading methods for linear dimensionality reduction include Principal Component Analysis (PCA), Factor Analysis (FA), Linear Multidimensional Scaling (MDS), Linear Discriminant Analysis (LDA), Canonical Correlation Analysis (CCA), Maximum Autocorrelation Factors (MAF), Slow Feature Analysis (SFA), Sufficient Dimensionality Reduction (SDR), Locality Preserving Projections (LPP), and Independent Component Analysis (ICA) \citep{cunninghamLinearDimensionalityReduction2015}. These approaches are important in single-cell transcriptomics for dissecting cellular heterogeneity, understanding cellular differentiation trajectories, and identifying correspondences between cells in different experiments \citep{xiang2021comparison, trapnell2014dynamics, stuartComprehensiveIntegrationSingleCell2019}.

\subsection{Unsupervised Methods for Linear Dimensionality Reduction}
\label{unsupervisedLDR}

Unsupervised linear dimensionality reduction methods, including PCA \citep{jolliffePrincipalComponentAnalysis2002}, ICA \citep{hyvarinenIndependentComponentAnalysis2001}, and FA \citep{spearmanGeneralIntelligenceObjectively1904}, project data into a low-dimensional space without the use of supervision labels. These methods are crucial in the initial stages of single-cell data analysis to reduce dimensionality and noise, and have been used in numerous studies to identify subpopulations of cells and understand the variance structure of the data \citep{xiang2021comparison, stuartComprehensiveIntegrationSingleCell2019}. The shortcoming of these unsupervised methods is that the axes of the low-dimensional space often fail to represent the underlying structure of the data, rendering them uninterpretable. This issue is particularly pronounced with gene expression data due to its high dimensionality (usually encompassing tens of thousands of genes) and the noisy expressions of many genes. These noisy expressions result in significant variance among individual cells, albeit without a structured pattern. In the absence of supervisory signals from phenotypic features, unsupervised methods tend to select these genes to construct the low-dimensional space, which does not provide the desired alignment or effective separation of cell type clusters. To illustrate this, we compared the performance of PCA on the gene expression data with that of FLDA. In brief, we solved the following objective to find the linear projection:

\begin{equation} \label{eq12}
\vu^{*} = \argmax_{\vu\in \mathbf{R}^{g}} \frac{\vu^T \mX \mX^{T} \vu}{\vu^T \vu}
\end{equation}

The results of this comparison are detailed in the Results section.

\subsection{Supervised Methods for Linear Dimensionality Reduction}
\label{supervisedLDR}

Supervised linear dimensionality reduction techniques, such as LDA \citep{fisherUseMultipleMeasurements1936, mclachlanDiscriminantAnalysisStatistical2004} and CCA \citep{hotellingRELATIONSTWOSETS1936, wangDeepMultiViewRepresentation2016}, can overcome the aforementioned issues. By incorporating supervised signals of phenotypic features, genes whose expressions do not inform on the phenotypes can be de-emphasized.

\subsubsection{Linear Discriminant Analysis (LDA)}
\label{linearDiscriminantAnalysis}

LDA models the differences among data organized in pre-determined classes. Formally, the optimization problem solved by LDA is as follows:

\begin{equation} \label{eq13}
\vu^{*} = \argmax_{\vu\in \mathbf{R}^{g}} \frac{\vu^T \mSigma_{b} \vu}{\vu^T \mSigma_{e} \vu}
\end{equation}

where $\mSigma_{b}$ and $\mSigma_{e}$ are estimates of the between-class and within-class covariance matrices, respectively.

Unlike FLDA, LDA doesn't explicitly formulate the representation of these classes as a contingency table composed of multiple features. As a result, when applied to an example problem where cell types are organized into a two-dimensional contingency table with phenotypic features $i$ and $j$, the axes from LDA are generally not aligned with these two phenotypic features.

However, it is possible to perform two separate LDAs for the two features. This modification allows the axes from each LDA to align with its specific feature. We refer to this approach as "2LDAs". There are two main limitations of this approach: first, it discards information about the component depending on the combination of the two features; second, it explicitly maximizes the segregation of cells with different feature levels, which sometimes is not consistent with a good separation of cell type clusters. Detailed comparisons between LDA, "2LDAs", and FLDA are provided in the Results section.

\subsubsection{Canonical Correlation Analysis (CCA)}
\label{canonicalCorrelationAnalysis}

CCA projects two datasets $\mX_{a} \in \mathbf{R}^{g \times n}$ and $\mX_{b} \in \mathbf{R}^{d \times n}$ to $\mY_{a} \in \mathbf{R}^{r \times n}$ and $\mY_{b} \in \mathbf{R}^{r \times n}$, such that the correlation between $\mY_{a}$ and $\mY_{b}$ is maximized. Formally, it tries to maximize this objective:

\begin{equation} \label{eq14}
(\vu^{}, \vv^{}) = \argmax_{{\vu\in \mathbf{R}^{g}}, \vv\in \mathbf{R}^{d}} \frac{\vu^T (\mX_{a} \mX_{a}^{T})^{-\frac{1}{2}} \mX_{a} \mX_{b}^{T} (\mX_{b} \mX_{b}^{T})^{-\frac{1}{2}} \vv}{(\vu^{T} \vu)^{-\frac{1}{2}} (\vv^{T} \vv)^{-\frac{1}{2}}}
\end{equation}

To apply CCA to our problem, we designate $\mX_{a}$ as the gene expression matrix, and $\mX_{b}$ as the matrix of $d$ phenotypic features ($d = 2$ for two features as demonstrated later). Unlike FLDA, CCA identifies a transformation of gene expressions that is aligned with a linear combination of phenotypic features, instead of a factorization of gene expressions corresponding to individual phenotypic features. The differences in these approaches are quantified and discussed in the Results section.

\subsection{Non-linear Dimensionality Reduction Methods}
\label{nonLinearDR}

Apart from linear dimensionality reduction, non-linear methods have emerged as popular choices for analyzing single-cell transcriptomic datasets due to their ability to capture complex, non-linear relationships inherent in the data \citep{xiang2021comparison}. Notable among these methods are t-Distributed Stochastic Neighbor Embedding (t-SNE) \citep{maaten2008visualizing} and Uniform Manifold Approximation and Projection (UMAP) \citep{mcinnes2018umap}. Unlike linear methods, these algorithms can unravel intricate structures in the data by modeling non-linear manifold structures.

t-SNE minimizes the divergence between two distributions over pairs of data points, one in the high-dimensional space and one in the low-dimensional space, to create a map that reflects the structure of the data. UMAP assumes that the data is uniformly distributed on a locally-connected Riemannian manifold and seeks to find a similar uniform distribution in lower dimensions.

The comparison of FLDA with t-SNE and UMAP hinges on the trade-off between linear and non-linear dimensionality reductions. While non-linear methods excel in capturing complex data structures and modeling dropout effects \citep{qiu2020embracing}, and often produce visually appealing embeddings, they exhibit certain limitations compared to linear methods, such as:

\begin{itemize}
    \item {Interpretability:} Linear methods offer a clear and direct relationship between the original features and the reduced dimensions, which facilitates interpretability. In contrast, the embeddings produced by non-linear methods are often challenging to interpret due to the complex and non-linear transformation functions involved.
    
    \item {Computational Efficiency:} Linear methods are generally more computationally efficient compared to non-linear methods, which can become computationally intensive, especially as the size of the dataset increases.
\end{itemize}

In single-cell transcriptomics applications, the choice between linear and non-linear dimensionality reduction hinges on balancing the capture of complex data structures with the maintenance of interpretability and computational efficiency. In the context of this paper, our proposed FLDA method is designed to address the challenges associated with single-cell data by offering a structured and interpretable low-dimensional space aligned with neuronal phenotypes. Therefore, we constrained our comparisons of FLDA with other linear dimensionality reduction methods that share the objective of interpretability.

\section{Experimental Design}
\label{experimentalDesign}

\subsection{Datasets}
\label{dataset}

\begin{figure}
\centering
\includegraphics[trim=0 400 0 100, clip,
width=0.95\linewidth]{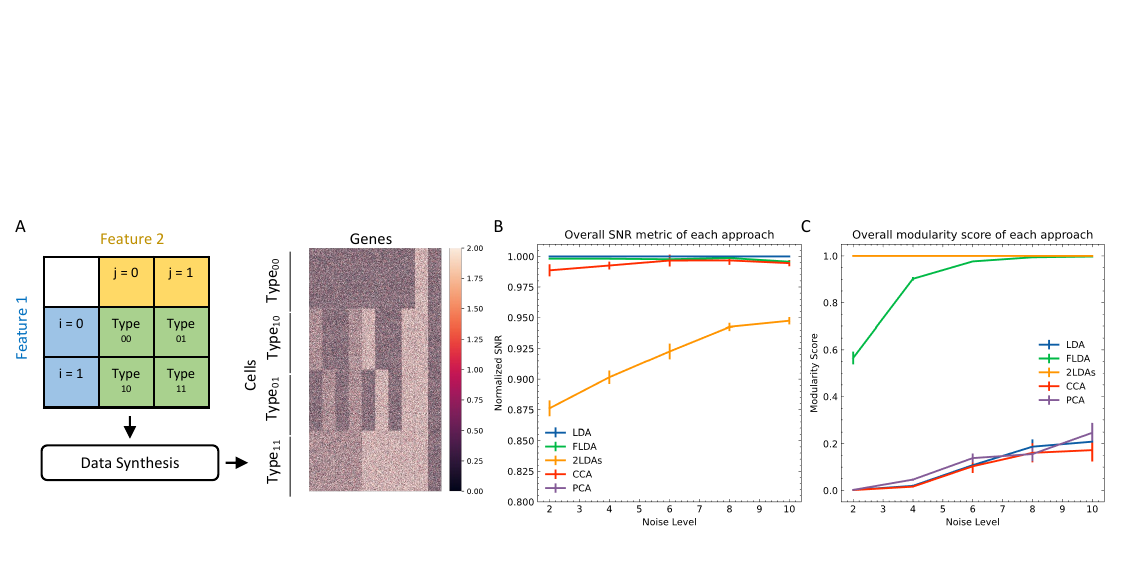}
\caption{Quantitative comparison between FLDA and other models. (A) Illustration of data synthesis. For implementation details, see Appendix C. The color bar represents the expression values of the 1000 generated genes. (B) Normalized overall Signal-to-Noise Ratio (SNR) metric for each analysis, normalized with respect to that of LDA. The normalized SNR metric of PCA is below 0.8. (C) Overall modularity score for each analysis. The error bars in (B,C) denote standard errors calculated from 10 repeated simulations. }
\label{fig2}
\end{figure}

\begin{figure}[hbt!]
\centering
\includegraphics[trim=10 350 0 85, clip, width=1\linewidth]{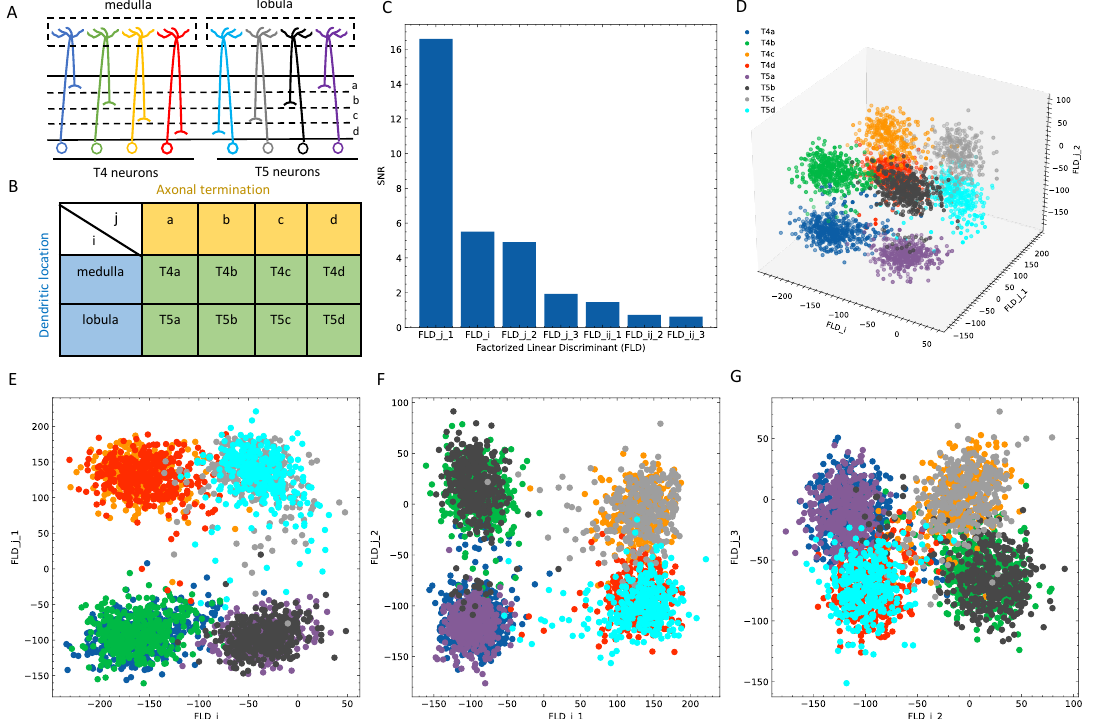}
\caption{Application of FLDA to the dataset of T4/T5 neurons. (A) T4/T5 neuronal cell types and their dendritic
and axonal location phenotypes. (B) The organization of T4/T5 neurons in a complete contingency table, where $i$ indicates dendritic location and $j$ indicates axonal termination. (C) SNR metric for each discriminant axis. (D) Data projection into the three-dimensional space consisting of the discriminant axis for feature $i$ (FLD$_i$) and the first and second discriminant axes for feature $j$ (FLD$_{j_1}$ and FLD$_{j_2}$). (E-G) Data projection into the two-dimensional space comprised of FLD$_i$ and FLD$_{j_1}$ (E), FLD$_{j_1}$ and FLD$_{j_2}$ (F), or FLD$_{j_2}$ and FLD$_{j_3}$ (the third discriminant axis for feature $j$) (G). Different cell types are represented by different colors as depicted in Panels (A) and (D).}
\label{fig3}
\end{figure}

To quantitatively evaluate FLDA against other linear dimensionality reduction methods such as PCA, CCA, LDA, and the "2LDAs" approach, we initially opted for synthetic datasets. The primary rationale behind this choice lay in the controlled environment synthetic data afford, enabling a precise and standardized comparison of the methods under varying conditions. These datasets consisted of four types of cells, each containing 250 examples, generated from a 2x2 Cartesian product of two features $i$ and $j$ (\Cref{fig2}A). We generated expressions for 1000 genes of each cell, with gene levels being either purely noise-driven or correlated with feature $i$, feature $j$, or the combination of both. Detailed information about the data generation can be found in Appendix C.

to bridge the gap between the controlled synthetic environment and real-world biological scenarios, we employed a dataset of Drosophila T4/T5 neurons \citep{kurmangaliyevModularTranscriptionalPrograms2019} to demonstrate the applicability and advantages of FLDA in analyzing single-cell transcriptome datasets. T4 and T5 neurons, while similar in general morphology and physiological properties, differ in the location of their dendrites in the medulla and lobula, which are two separate brain regions. Both T4 and T5 neurons comprise four subtypes, each pair demonstrating axonal lamination in a specific layer within the lobula plate (\Cref{fig3}A). Thus, we identified these neurons using two phenotypic features: feature $i$ indicating the dendritic location in either the medulla or lobula, and feature $j$ signifying axonal lamination at one of the four layers (a/b/c/d) (\Cref{fig3}B). In this study, we concentrated on a dataset containing expression data for 17492 genes from 3833 cells, all collected at a predefined time during brain development.

\subsection{Data Preprocessing}
\label{preprocessing}

The preprocessing of the T4/T5 neuron dataset adhered to previously documented procedures \citep{shekharCOMPREHENSIVECLASSIFICATIONRETINAL2016, pengMolecularClassificationComparative2019, tranSinglecellProfilesRetinal2019, kurmangaliyevModularTranscriptionalPrograms2019}. Briefly, the transcript counts within each column of the count matrix (genes$\times$cells) were normalized to equate to the median number of transcripts per cell, leading to normalized counts, or Transcripts-per-million ($TPM_{gc}$), for Gene $g$ in Cell $c$. We used the log-transformed expression data, denoted by $E_{gc} = \ln {(TPM_{gc}+1)}$, for subsequent analysis. We selected highly variable genes for further FLDA application based on a common approach in single-cell RNA-Seq studies. This approach is based on establishing a relationship between mean and coefficient of variation \citep{chenDetectionHighVariability2016, pandeyComprehensiveIdentificationSpatial2018, kurmangaliyevModularTranscriptionalPrograms2019}. For this particular experiment, we set the hyper-parameters $\lambda$s in \Cref{eq2} to 1.

\subsection{Evaluation Metrics}
\label{metrics}

The dimensionality reduction process should satisfy two primary goals: (1) to identify axes that efficiently segregate distinct cell types, and (2) to discover axes that are well-aligned with the respective labels. Consequently, to evaluate the effectiveness of FLDA and various alternative methodologies, we implemented the following metrics (Detailed information of implementing these metrics can be found in Appendix D):

\begin{itemize}
\item Signal-to-Noise Ratio (SNR): This metric measures the efficacy of each discriminant axis in distinguishing distinct cell types. Higher SNR suggests better separation of different cell types. This metric is relevant to the first goal.
\item Explained Variance (EV): This metric gauges the proportion of variance of the feature $i$ or $j$ that a discriminant axis explains. Higher EV indicates that the dimensionality reduction method effectively encapsulates the feature information. This metric is relevant to the second goal.
\item Mutual Information (MI): This metric calculates the association between each discriminant axis and each feature, providing insights into how much information an axis provides about a specific feature. A higher MI score suggests better ability of the dimensionality reduction method to capture essential characteristics. This metric is relevant to the second goal.
\item Modularity Score: This metric assesses whether each axis is predominantly dependent on a single feature \citep{ridgewayLearningDeepDisentangled2018}. A higher modularity score indicates successful disentanglement of features, which is crucial for interpreting biological data. This metric is relevant to the second goal.
\item Silhouette Score: This metric computes the average Silhouette value of all samples, which is a measure of how similar a cell is to its own cluster compared to other clusters. A higher Silhouette score indicates better cluster separation and tighter clustering, This metric is relevant to the first goal.
\end{itemize}

In addition, we evaluated the execution times of FLDA and alternative methodologies.

\section{Results}
\label{results}

\subsection{Comparative Analysis of FLDA with Other Linear Dimensionality Reduction Methods}
\label{modelComparisons}

To provide a quantitative comparison between FLDA and other dimensionality reduction methods such as PCA, CCA, LDA, and "2LDAs", we measured the proposed metrics on the synthesized datasets as shown in \Cref{fig2}A. Given that the synthesized data was organized in a 2x2 table, each LDA of the "2LDAs" approach could only identify one dimension for the specific features $i$ or $j$. Therefore, as a fair comparison, we only included the corresponding dimensions in FLDA (FLD$_i$ and FLD$_j$) and the top two components of PCA, CCA, and LDA. The overall SNR values normalized by that of LDA and the modularity scores across different noise levels are depicted in \Cref{fig2}B,C. The performance of PCA is the worst due to its unsupervised approach, which cannot effectively mitigate the impact of noise on the signal. While supervised approaches generally demonstrate superior SNR, LDA and CCA suffer from low modularity scores. This outcome aligns with our expectation, as LDA maximizes cell type cluster separation without necessarily aligning axes to individual features $i$ or $j$, and CCA maximizes the correlation to a linear combination of phenotypic features rather than individual ones. Conversely, "2LDAs" achieves the highest modularity scores but exhibits the lowest SNR among supervised approaches, as it aims to maximize the separation of cells with different feature levels, which does not necessarily coincide with maximizing cell type segregation. Both the SNR and modularity score of FLDA approach optimal values because it considers both the alignment of axes to different features and the constraint of variance within cell types. Consistent with the SNR metric, the average Silhouette score for FLDA is close to those of LDA and CCA, outperforms "2LDAs", and significantly surpasses PCA, as detailed in \Cref{tab:table1}. Consistent with the modularity score, a robust axis alignment to either feature $i$ or $j$ is observed in FLDA and "2LDAs", but not in the other methods, as shown in a representative plot of the EV and MI metrics across these models in \Cref{AFig1}. 

We further analyzed the execution times of FLDA and other models and summarized the findings in \Cref{tab:table2}. The execution time of FLDA is on par with that of LDA, albeit longer than PCA's, attributed to the handling of the covariance matrix in the denominator. In contrast, the execution times for "2LDAs" and CCA are considerably extended, nearly doubling those of FLDA and LDA. This increment is due to "2LDAs" requiring two LDA operations, while CCA necessitates the computation of covariance matrices for both input and phenotypic features, thereby doubling the execution time.

\subsection{Real-world Application in Computational Biology}
\label{application}

A significant question in biology is whether diverse cell type phenotypes are generated by modular transcriptional programs, and if so, what the gene signature for each program is. To demonstrate the potential of our approach in addressing this question, we applied FLDA to the Drosophila T4/T5 neuron dataset. 

Given that the data is organized in a 2x4 contingency table, we chose to project the expression data into a seven-dimensional subspace. This subspace was structured such that one FLD was aligned with dendritic location $i$ (FLD$_i$), three FLDs were aligned with axonal termination $j$ (FLD${_{j_{1-3}}}$), and the remaining three were tailored to represent the combination between both phenotypes (FLD${_{{ij}_{1-3}}}$). Ranking these axes based on their SNR metrics revealed that FLD$_{j_1}$, FLD$_i$, and FLD$_{j_2}$ had considerably higher SNRs than the others (\Cref{fig3}C). Indeed, data representations in the subspace comprising these three dimensions clearly separated the eight neuronal cell types (\Cref{fig3}D). As expected, FLD$_i$ differentiated T4 from T5 neurons, which have dendrites located in different brain regions (\Cref{fig3}E). Interestingly, FLD$_{j_1}$ separated T4/T5 neurons into two groups, a/b vs c/d, according to the upper or lower lobula place, while FLD$_{j_2}$ divided them into another two groups, a/d vs b/c, indicating whether their axons laminated at the middle or lateral part of the lobula plate (\Cref{fig3}E,F). Among these three dimensions, FLD$_{j_1}$ has a much higher SNR than FLD$_i$ and FLD$_{j_2}$, suggesting a hierarchical structure in the genetic organization of T4/T5 neurons: they are first separated into either a/b or c/d types, and subsequently divided into each of the eight subtypes. In fact, this matches the sequence of their cell fate determination, as revealed in a previous genetic study \citep{pinto-teixeiraDevelopmentConcurrentRetinotopic2018}. Lastly, the final discriminant axis of the axonal feature FLD$_{j_3}$ separates the group a/c from b/d, suggesting its role in fine-tuning the axonal depth within the upper or lower lobula plate (\Cref{fig3}G).

\begin{figure}
  \centering
  \includegraphics[trim=10 455 80 85, clip, width=0.975\linewidth]{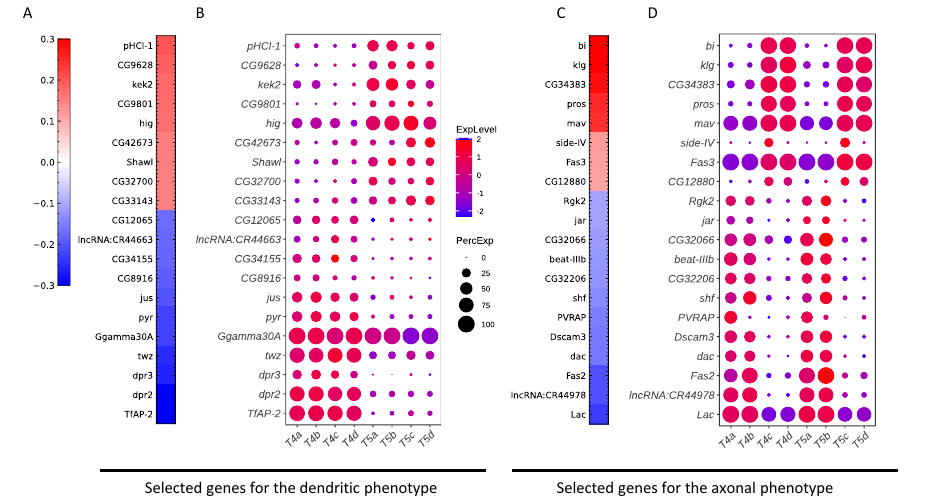}
  \caption{Critical genes extracted from the sparse algorithm. (A) Weight vector of the 20 genes selected for the dendritic phenotype (FLD$_i$). The weight value is indicated in the color bar with color indicating direction (red: positive and green: negative) and saturation indicating magnitude. (B) Expression patterns of the 20 genes from (A) in eight types of T4/T5 neurons. Dot size indicates the percentage of cells in which the gene was expressed, and color represents average scaled expression. (C) Weight vector of the 20 genes selected for the axonal phenotype (FLD$_{j_1}$). Legend as in (A). (D) Expression patterns of the 20 genes from (C) in eight types of T4/T5 neurons. Legend as in (B).}
  \label{fig4}
\end{figure}

\begin{figure}[hbt!]
  \centering
  \includegraphics[trim=0 500 100 85, clip, width=1\linewidth]{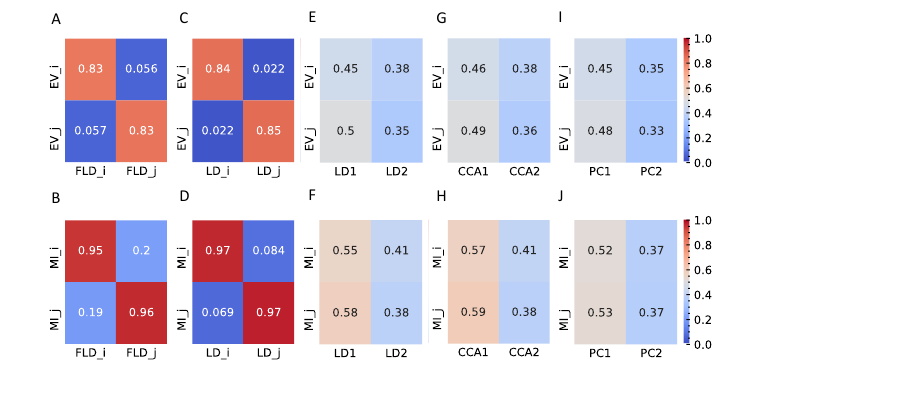}
  \caption{Representative plots (at $\sigma = 6$) of EV and MI metrics for FLDA and other models. (A,B) EV (A) and MI (B) metrics of FLDA. FLD$_{i}$ and FLD$_{j}$ indicate the factorized linear discriminants for features $i$ and $j$. (C,D) EV (C) and MI (D) metrics of 2LDAs. LD$_{i}$ and LD$_{j}$ indicate the linear discriminant components for features $i$ and $j$. (E,F) EV (E) and MI (F) metrics of LDA. LD$_{1}$ and LD$_{2}$ indicate the first two linear discriminant components. (G,H) EV (G) and MI (H) metrics of CCA. CCA$_{1}$ and CCA$_{2}$ indicate the first two canonical correlation axes. (I,J) EV (I) and MI (J) metrics of PCA. PC$_{1}$ and PC$_{2}$ indicate the first two principal components. EV$_{i}$ and EV$_{j}$ are the explained variance of features $i$ and $j$ along an axis, and MI$_{i}$ and MI$_{j}$ indicate the mutual information between an axis and features $i$ and $j$ respectively. Values of EV and MI metrics are also indicated by the color bars on the right side. }
  \label{AFig1}
\end{figure}

\begin{figure}[hbt!]
  \centering
  \includegraphics[trim=0 525 100 85, clip, width=1\linewidth]{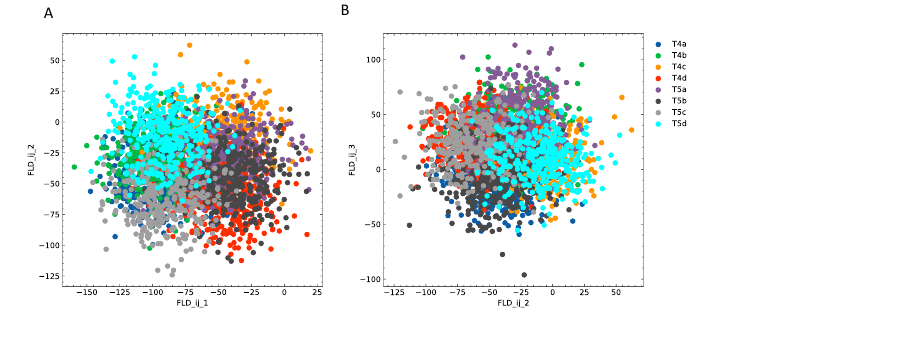}
  \caption{Additional plots for FLDA on the dataset of T4/T5 neurons. (A, B) Projection of the original gene expression data into the two-dimensional space made of the first and second (FLD$_{{ij}_1}$ and FLD$_{{ij}_2}$) (A) or the second and third (FLD$_{{ij}_2}$ and FLD$_{{ij}_3}$) (B) discriminant axes for the component that depends on the combination of both features $i$ and $j$. Different cell types are indicated in different colors as in panel B.}
  \label{AFig2}
\end{figure}

\begin{figure}[hbt!]
  \centering
  \includegraphics[trim=10 475 80 85, clip, width=0.975\linewidth]{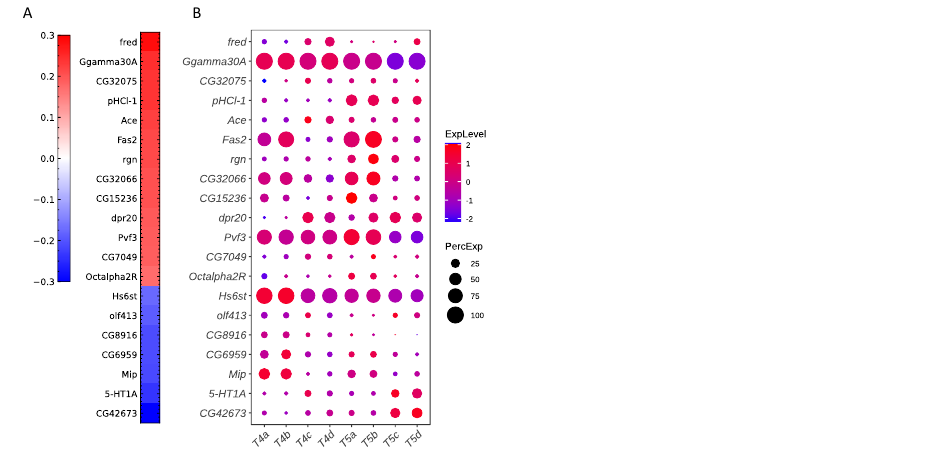}
  \caption{Additional plots for critical genes extracted from the sparse algorithm. (A) Weight vector of the 20 genes selected for the combination of dendritic and axonal features (FLD$_{{ij}_1}$). The weight value is indicated in the color bar with color indicating direction (red: positive and green: negative) and saturation indicating magnitude. (B) Expression patterns of the 20 genes from (A) in eight types of T4/T5 neurons. Dot size indicates the percentage of cells in which the gene was expressed, and color represents average scaled expression.}
  \label{AFig3}
\end{figure}

To identify gene signatures for the discriminant components in FLDA, we applied sparsity-based regularization to constrain the number of genes with non-zero weight coefficients. We set the number to 20, a reasonable number of candidate genes that could be tested in a follow-up biological study. We extracted a list of 20 genes each for the axis of FLD$_i$ or FLD$_{j_1}$. The relative importance of these genes to each axis is directly informed by their weight values (\Cref{fig4}A,C). Alongside, we plotted expression profiles of these genes in the eight neuronal cell types (\Cref{fig4}B,D). For both axes, the genes critical in separating cells with different feature levels are differentially expressed in corresponding cell types. Finally, FLDA allowed us to examine the component that depends on the combination of both features and identify its gene signature, providing insights into transcriptional regulation of gene expressions in the T4/T5 neuronal cell types (\Cref{AFig2,AFig3}).

\subsection{Perturbation Analysis}
\label{perturbation}

\begin{figure}[hbt!]
  \centering
  \includegraphics[trim=0 425 100 85, clip, width=1\linewidth]{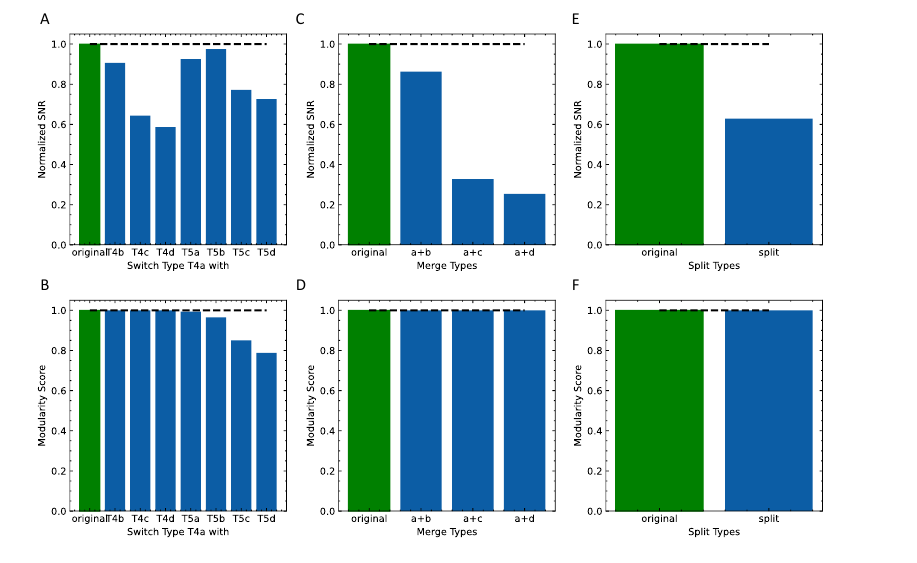}
  \caption{Evaluation of the effect of incorrect phenotype annotation on the dataset of T4/T5 neurons. (A,B) Normalized overall SNR metric (A) and overall modularity score (B) of FLDA after switching labels of T4a type with another neuronal type. (C,D) Normalized overall SNR metric (C) and overall modularity score (D) of FLDA after merging the axonal phenotypic level a with another phenotypic level (b/c/d). (E,F) Normalized overall SNR metric (E) and overall modularity score (F) of FLDA after splitting each axonal phenotypic level into two. Metrics under the original annotation are colored in green, and their values are indicated by the dashed lines. Here the SNR values are normalized with respect to that of the original annotation.}
  \label{AFig4}
\end{figure}

As FLDA, like other supervised methods, relies on accurate phenotype labels to extract meaningful information, we sought to investigate how it might behave in real-world scenarios where inaccuracies are bound to occur. If the phenotypes are annotated incorrectly, can we use FLDA to raise a flag? To address this, we propose a perturbation analysis of FLDA, based on the assumption that among possible phenotype annotations, the projection of gene expression data with correct labels leads to better metric measurements than incorrect ones. As detailed in Appendix E, we deliberately generated three kinds of incorrect labels for the T4/T5 neuron dataset, simulating common erros that could occur during labeling: the phenotypes of a cell type were mislabeled with those of another type; a singular phenotypic category was incorrectly split into two; two phenotypic categories were incorrectly merged into one. We applied FLDA to gene expressions of T4/T5 neurons using these perturbed annotations, and found that proposed metrics, such as SNR and the modularity score, were best when the labels were correct (\Cref{AFig4}), suggesting that this type of perturbation analysis can be used to flag potential errors in labeling.

In summary, our findings demonstrate that FLDA is a powerful tool for identifying and interpreting gene expressions that correspond to particular phenotypic features, even in the face of potential data mislabeling. This makes it a valuable tool for understanding complex biological systems. The perturbation analysis provides a robust method for validating the accuracy of phenotype annotations, thereby increasing the reliability of subsequent analyses and conclusions.

\section{Discussion}
\label{discussion}

We have introduced FLDA, a novel dimensionality reduction method that linearly projects high-dimensional data, such as gene expressions, into a low-dimensional space. The axes of this space are aligned with predefined features like phenotypes, making it an intuitive representation. Furthermore, we incorporated sparse regularization into FLDA, allowing us to select a small set of critical genes that are most informative about the phenotypes. Our application of FLDA in a computational biology context, particularly in the analysis of gene expression data from Drosophila T4/T5 neurons with two phenotypic labels, not only illuminated data structures aligned with the phenotypic labels, but also unveiled previously unreported genes associated with each phenotype. A comparison of our gene lists with those from the previous study \citep{kurmangaliyevModularTranscriptionalPrograms2019} unveiled consistent genes including indicator genes for dendritic location like $TfAP-2$, $dpr2$, $dpr3$, $twz$, $CG34155$, and $CG12065$, and those for axonal lamination such as $klg$, $bi$, $pros$, $mav$, $beat-IIIb$, and $Fas2$. Remarkably, we identified genes not reported in the previous study. For example, our results suggest that the gene $pHCl-1$ is important to the dendritic phenotype, and the gene $Lac$ is critical to axonal lamination. These genes are promising genetic targets for subsequent experimentation.

FLDA's potential extends beyond the dataset explored in this study. In a separate work, we applied FLDA to another real-world single-cell transcriptomic dataset, showcasing its ability to discern a low-dimensional representation of neuronal types aligned with phenotypic and species attributes, thereby revealing evolutionary counterparts of primate retinal ganglion cells \citep{hahn2023evolution}. This further substantiates FLDA's applicability across diverse datasets and its promise in unveiling biologically meaningful insights.

The method could also play a role in the discovery of cell types. For example, the known phenotypes in a population might only form a partial table with missing entries (Figure 1B). Like the empty cells in Mendeleev's Periodic Table led to the prediction of new elements, these gaps could indicate predictions of new cell types \citep{mendelejeffUeberBeziehungenEigenschaften1869}. FLDA can help pinpoint the region of the gene expression space that corresponds to the predicted new type, potentially revealing rare cell populations that might otherwise be overlooked due to insignificance.

Beyond computational biology, FLDA's application can extend to any labeled dataset with labels forming a Cartesian product of multiple attributes. This ability to separate attribute-specific factors makes FLDA invaluable in creating disentangled representations \citep{ridgewayLearningDeepDisentangled2018, karaletsosBayesianRepresentationLearning2016}. The potential of FLDA extends to these areas, and its performance can be optimized for diverse applications.

While our work offers significant advancements, it is not without limitations. The inherent linearity of FLDA, though providing an explicit and easily interpretable model, also presupposes a linear relationship between input features, which may not always hold true. Future work could involve a non-linear version of FLDA. For example, the input features can be projected into an embedding space using a neural network, where the axes align with each label attribute.

\section{Appendix}
\label{appendix}

\subsection{A. Objective functions}
\label{appendixObjectiveFunctions}

Here we derive the objective functions used in our analysis. Again if $\vx_{ijk} (k \in {1,2,...n_{ij}})$ represents the expression values of $g$ genes in each cell ($\vx_{ijk} \in \mathbf{R}^{g})$), we seek to find a linear projection $y_{ijk} = \vu^T \vx_{ijk}$ that is aligned with the feature $i$.

\subsubsection{Inspiration from ANOVA}
\label{appendixObjectiveFunctionsANOVA}

We asked what is the best way to factorize $y_{ijk}$. Inspired by multi-way ANOVA \citep{fisherCorrelationRelativesSupposition1918}, we identified three components: one depending on the feature $i$, another depending on the feature $j$, and the last one depending on the combination of both features. We therefore followed the procedures of ANOVA to partition sums of squares and factorize $y_{ijk}$ into these three components.

Let us first assume that all cell types defined by $i$ and $j$ contain the same number of cells. With cell types represented by a complete contingency table (\Cref{fig1}A), $y_{ijk}$ can be linearly factorized using the model of two crossed factors. Formally, the linear factorization is the following:

\begin{equation} \label{eqA0}
y_{ijk} = \mu + \alpha_{i} + \beta_{j} + (\alpha\beta)_{ij}+\epsilon_{ijk}
\end{equation}

where $y_{ijk}$ represents the coordinate of the $k$th cell in the category defined by $i$ and $j$; $\mu$ is the average level of $y$; $\alpha_i$ is the component that depends on the feature $i$, and $\beta_j$ is the component that depends on the feature $j$; $(\alpha\beta)_{ij}$ describes the component that depends on the combination of both features $i$ and $j$; $\epsilon_{ijk} \sim \mathcal{N} (0,\sigma^2)$ is the residual of this factorization.

Let us say that the features $i$ and $j$ fall into $a$ and $b$ discrete categories respectively. Then without loss of generality, we can require:

\begin{equation} \label{eqA1}
\sum_{i=1}^{a} \alpha_i = 0
\end{equation}
	
\begin{equation} \label{eqA2}
\sum_{j=1}^{b} \beta_j = 0
\end{equation}

\begin{equation} \label{eqA3}
\sum_{i=1}^{a} (\alpha\beta)_{ij} = \sum_{j=1}^{b} (\alpha\beta)_{ij} = 0
\end{equation}
	
Corresponding to these, there are three null hypotheses:

\begin{equation} \label{eqA4}
H_{01}: \alpha_i = 0
\end{equation}

\begin{equation} \label{eqA5}
H_{02}: \beta_j = 0
\end{equation}

\begin{equation} \label{eqA6}
H_{03}: (\alpha\beta)_{ij} = 0
\end{equation}

Here we want to reject $H_{01}$ while accepting $H_{02}$ and $H_{03}$ so that $y_{ijk}$ is aligned with the feature $i$.

Next, we partition the total sum of squares. If the number of cells within each cell type category is $n$, and the total number of cells is $N$, then we have

\begin{equation} \label{eqA7}
\begin{split} 
\sum_{i=1}^{a} \sum_{j=1}^{b} \sum_{k=1}^{n} (y_{ijk} - \bar{y}_{...})^2
&= bn\sum_{i=1}^{a}(\bar{y}_{i..} - \bar{y}_{...})^2
+ an\sum_{j=1}^{b}(\bar{y}_{.j.} - \bar{y}_{...})^2
\\
&+ n\sum_{i=1}^{a} \sum_{j=1}^{b} (\bar{y}_{ij.} - \bar{y}_{i..} - \bar{y}_{.j.} + \bar{y}_{...})^2
+ \sum_{i=1}^{a} \sum_{j=1}^{b} \sum_{k=1}^{n} (y_{ijk} - \bar{y}_{ij.})^2
\end{split}
\end{equation}

where $\bar{y}$ is the average of $y_{ijk}$ over the indices indicated by the dots. \Cref{eqA7} can be written as

\begin{equation} \label{eqA8}
SS_{T} = SS_{A} + SS_{B} + SS_{AB} + SS_{e}
\end{equation}

with each term having degrees of freedom $N-1$, $a-1$, $b-1$, $(a-1)(b-1)$, and $N-ab$ respectively. Here $SS_{A}$, $SS_{B}$, $SS_{AB}$, and $SS_{e}$ are partitioned sum of squares for the factors $\alpha_i$, $\beta_j$, $(\alpha\beta)_{ij}$, and the residual.

ANOVA rejects or accepts a null hypothesis by comparing its mean square (the partitioned sum of squares normalized by the degree of freedom) to that of the residual. This is done by constructing F-statistics for each factor as shown below:

\begin{equation} \label{eqA9}
F_{A} = \frac{MS_{A}}{MS_{e}} = \frac{\frac{SS_{A}}{a-1}}{\frac{SS_{e}}{N-ab}}
\end{equation}

\begin{equation} \label{eqA10}
F_{B} = \frac{MS_{B}}{MS_{e}} = \frac{\frac{SS_{B}}{b-1}}{\frac{SS_{e}}{N-ab}}
\end{equation}

\begin{equation} \label{eqA11}
F_{AB} = \frac{MS_{AB}}{MS_{e}} = \frac{\frac{SS_{AB}}{(a-1)(b-1)}}{\frac{SS_{e}}{N-ab}}
\end{equation}

Under the null hypotheses, the F-statistics follow the F-distribution. Therefore, a null hypothesis is rejected when we observe the value of a F-statistic above a certain threshold calculated from the F-distribution. Here we want $F_{A}$ to be large enough so that we can reject $H_{01}$, but $F_{B}$ and $F_{AB}$ to be small enough for us to accept $H_{02}$ and $H_{03}$. In other words, we want to maximize $F_{A}$ while minimizing $F_{B}$ and $F_{AB}$. Therefore, we propose maximizing an objective $L$:

\begin{equation} \label{eqA12}
L = F_{A} - \lambda_{1} F_{B} - \lambda_{2} F_{AB}
\end{equation}

where $\lambda_{1}$ and $\lambda_{2}$ are hyper-parameters determining the relative weights of $F_{B}$ and $F_{AB}$ compared with $F_{A}$.

\subsubsection{Objective functions under a complete contingency table}
\label{appendixObjectiveFunctionsComplete}

When the numbers of cells within categories defined by $i$ and $j$ ($n_{ij}$) are not all the same, the total sum of squares cannot be partitioned as in \Cref{eqA7}. However, if we only care about distinctions between cell types instead of individual cells, we can use the mean value of each cell type cluster ($\bar{y}_{ij.}$) to estimate the overall average value ($\tilde{y}_{...}$), and the average value of each category $i$ ($\tilde{y}_{i..}$) or $j$ ($\tilde{y}_{.j.}$). Therefore, \Cref{eqA7} can be modified as the following:

\begin{equation} \label{eqA13}
\begin{split} 
\sum_{i=1}^{a} \sum_{j=1}^{b} [\frac{1}{n_{ij}} \sum_{k=1}^{n_{ij}} (y_{ijk} - \tilde{y}_{...})^2]
&= b \sum_{i=1}^{a}(\tilde{y}_{i..} - \tilde{y}_{...})^2
+ a \sum_{j=1}^{b}(\tilde{y}_{.j.} - \tilde{y}_{...})^2
\\
&+ \sum_{i=1}^{a} \sum_{j=1}^{b} (\bar{y}_{ij.} - \tilde{y}_{i..} - \tilde{y}_{.j.} + \tilde{y}_{...})^2
+ \sum_{i=1}^{a} \sum_{j=1}^{b} [\frac{1}{n_{ij}} \sum_{k=1}^{n_{ij}} (y_{ijk} - \bar{y}_{ij.})^2]
\end{split}
\end{equation}

where

\begin{equation} \label{eqA14}
\bar{y}_{ij.} = \frac{\sum_{k=1}^{n_{ij}} y_{ijk}}{n_{ij}}
\end{equation}

\begin{equation} \label{eqA15}
\tilde{y}_{i..} = \frac{\sum_{j=1}^{b} \bar{y}_{ij.}}{b}
\end{equation}

\begin{equation} \label{eqA16}
\tilde{y}_{.j.} = \frac{\sum_{i=1}^{a} \bar{y}_{ij.}}{a}
\end{equation}

\begin{equation} \label{eqA17}
\tilde{y}_{...} = \frac{\sum_{i=1}^{a} \sum_{j=1}^{b} \bar{y}_{ij.}}{ab}
\end{equation}

If we describe \Cref{eqA13} as:

\begin{equation} \label{eqA18}
\tilde{SS}_{T} = \tilde{SS}_{A} + \tilde{SS}_{B} + \tilde{SS}_{AB} + \tilde{SS}_{e}
\end{equation}

then following the same arguments, we want to maximize an objective function in the following format:

\begin{equation} \label{eqA19}
L = \frac{\frac{\tilde{SS}_{A}}{a-1} - \lambda_{1} \frac{\tilde{SS}_{B}}{b-1} - \lambda_{2} \frac{\tilde{SS_{AB}}}{(a-1)(b-1)} }{\frac{\tilde{SS}_{e}}{N-ab}}
\end{equation}

\subsubsection{Objective functions under a partial contingency table}
\label{appendixObjectiveFunctionsPartial}

When we have a representation of a partial table, we can no longer separate out the component that depends on the combination of both features. Therefore, we use another model, a linear model of two nested factors, to factorize $y_{ijk}$, which has the following format:

\begin{equation} \label{eqA20}
y_{ijk} = \mu + \alpha_{i} + \beta_{j(i)} + \epsilon_{ijk}
\end{equation}

Note that we now have $\beta_{j(i)}$ instead of $\beta_j + (\alpha\beta)_{ij}$. In this model, we identify a primary factor, for instance, the feature denoted by $i$ which falls into $a$ categories, and the other (indexed by $j$) becomes a secondary factor, the number of whose levels $b_{i}$ depends on the level of the primary factor. We merge the component depending on the combination of both features into that of the secondary factor as $\beta_{j(i)}$.

Similarly, we have

\begin{equation} \label{eqA21}
\begin{split}
\sum_{i=1}^{a} \sum_{j =1}^{b_{i}} [\frac{1}{n_{ij}}\sum_{k=1}^{n_{ij}} (y_{ijk} - \tilde{y}_{...})^2]
&= \sum_{i=1}^{a}[\sum_{j=1}^{b_{i}} (\tilde{y}_{i..} - \tilde{y}_{...})^2] \\ 
&+ \sum_{i=1}^{a} \sum_{j=1}^{b_{i}} (\bar{y}_{ij.} - \tilde{y}_{i..})^2
+ \sum_{i=1}^{a} \sum_{j=1}^{b_{i}} [\frac{1}{n_{ij}} \sum_{k=1}^{n_{ij}} (y_{ijk} - \bar{y}_{ij.})^2]
\end{split}
\end{equation}

which can be written as

\begin{equation} \label{eqA22}
\tilde{SS}_{T} = \tilde{SS}_{A} + \tilde{SS}_{B} + \tilde{SS}_{e}
\end{equation}

with degrees of freedom $N-1$, $a-1$, $M-a$, and $N-M$ for each of the terms, where $M$ is:

\begin{equation} \label{eqA23}
M = \sum_{i=1}^{a} b_{i}
\end{equation}

Therefore, we want to maximize the following objective:

\begin{equation} \label{eqA24}
L = \frac{\frac{\tilde{SS}_{A}}{a-1} - \lambda \frac{\tilde{SS}_{B}}{M-a}}{\frac{\tilde{SS}_{e}}{N-M}}
\end{equation}

\subsection{B. FLDA with a partial contingency table}
\label{appendixFLDAPartialTable}

Here we provide the mathematical details of FLDA under the representation of a partial table. When we have a partial table, if the feature $i$ is the primary feature with $a$ levels, and the feature $j$ is the secondary feature with $b_{i}$ levels, then $\mN_{A}$ in \Cref{eq1} is defined as follows:

\begin{equation} \label{eqA25}
\mN_{A} =  \mM_{A} - \lambda \mM_{B|A}
\end{equation}

where

\begin{equation} \label{eqA26}
\mM_{A} = \frac{1}{a-1}\sum_{i=1}^{a}\sum_{j=1}^{b_{i}} (\vm_{i.} - \vm_{..})(\vm_{i.} - \vm_{..})^T
\end{equation}

\begin{equation} \label{eqA27}
\mM_{B|A} = \frac{1}{M-a}\sum_{i=1}^{a}\sum_{j=1}^{b_{i}} (\vm_{ij} - \vm_{i.})(\vm_{ij} - \vm_{i.})^T
\end{equation}

and $M$ is defined as in \Cref{eqA23}. Correspondingly, $\mM_{e}$ in \Cref{eq1} is defined as:

\begin{equation} \label{eqA28}
\mM_{e} = \frac{1}{N-M}\sum_{i=1}^{a}\sum_{j=1}^{b}[\frac{1}{n_{ij}}\sum_{k=1}^{n_{ij}} (\vx_{ijk} - \vm_{ij})(\vx_{ijk} - \vm_{ij})^T]
\end{equation}

and

\begin{equation} \label{eqA29}
\vm_{..} = \frac{1}{M}\sum_{i=1}^{a} \sum_{j=1}^{b_{i}} \vm_{ij}
\end{equation}

\begin{equation} \label{eqA30}
\vm_{i.} = \frac{1}{b_{i}}\sum_{j=1}^{b_{i}} \vm_{ij}
\end{equation}

The remaining mathematical arguments are the same as those for the complete table. In this scenario, because we don't observe all possible combinations of features $i$ and $j$, we cannot find the linear projection for the combination of both features.

\subsection{C. Implementation details of data synthesis}
\label{appendixImplementationOfDataSynthesis}

To quantitatively compare FLDA with alternative approaches, we synthesized data of four cell types, each of which contained 250 cells. The four cell types were generated from a Cartesian product of two features $i$ and $j$, where $i \in \{0,1\}$ and $j \in \{0,1\}$. Expressions of 1000 genes were generated for each cell. The expression value of the $h^{\rm{th}}$ gene in the $k^{\rm{th}}$ cell of the cell type $ij$, $x_{ijk}^{h}$ was defined as the following:

\begin{equation} \label{eqA31}
x_{ijk}^{1:100} = i+\epsilon_{ijk}
\end{equation}

\begin{equation} \label{eqA32}
x_{ijk}^{101:200} = j+\epsilon_{ijk}
\end{equation}

\begin{equation} \label{eqA33}
x_{ijk}^{201:300} = i \land j+\epsilon_{ijk}
\end{equation}

\begin{equation} \label{eqA34}
x_{ijk}^{301:400} = i \lor j+\epsilon_{ijk}
\end{equation}

\begin{equation} \label{eqA35}
x_{ijk}^{401:500} = 2i+\epsilon_{ijk}
\end{equation}

\begin{equation} \label{eqA36}
x_{ijk}^{501:600} = 2j+\epsilon_{ijk}
\end{equation}

\begin{equation} \label{eqA37}
x_{ijk}^{601:700} = 2i \land j+\epsilon_{ijk}
\end{equation}

\begin{equation} \label{eqA38}
x_{ijk}^{701:800} = 2i \lor j+\epsilon_{ijk}
\end{equation}

\begin{equation} \label{eqA39}
x_{ijk}^{801:900} = \epsilon_{ijk}
\end{equation}

\begin{equation} \label{eqA40}
x_{ijk}^{901:1000} = 2 + \epsilon_{ijk}
\end{equation}
where

\begin{equation} \label{eqA41}
    i \land j= 
\begin{cases}
    1,      & \text{if } i = 1, j = 1\\
    0,               & \text{otherwise}
\end{cases}
\end{equation}
and 

\begin{equation} \label{eqA42}
    i \lor j= 
\begin{cases}
    0,      & \text{if } i = 0, j = 0\\
    1,               & \text{otherwise}
\end{cases}
\end{equation}
were combinations of the two features. Here $\epsilon_{ijk}$ represents Gaussian noise, namely

\begin{equation} \label{eqA43}
\epsilon_{ijk} \sim \mathcal{N} (0,\sigma^2)
\end{equation}
We generated synthetic datasets under varying levels of Gaussian noise by adjusting the $\sigma$ value across a set of 5 distinct levels ($\sigma \in {2, 4, 6, 8, 10}$). For each of these $\sigma$ values, we created 10 distinct datasets in a repetitive manner, leading to a total of 50 synthetic datasets. The evaluation metrics for each $\sigma$ value were then computed as the average across the respective 10 repetitions, providing a robust evaluation of the proposed FLDA method across different noise levels.

\subsection{D. Implementation details of the metrics used in the study}
\label{appendixImplementationOfMetrics}

We measured the following metrics in our experiments:

\subsubsection{Signal-to-Noise Ratio (SNR)}
\label{appendixSNR}

Because we care about the separation of cell types, we define the SNR metric as the ratio of the variance between cell types over the variance of the noise, which is estimated from within-cluster variance. For the entire embedding space, given $q$ cell types, if the coordinate of each cell is indicated by $\vx$, then we define the overall SNR metric as the following:

\begin{equation} \label{eqA44}
SNR_{overall} = \frac{tr(\Sigma_{p=1}^{q}n_{p}(\bar{\vx}_{p.} - \bar{\vx}_{..})(\bar{\vx}_{p.} - \bar{\vx}_{..})^{T}))}{tr(\Sigma_{p=1}^{q}\Sigma_{k=1}^{n_{p}}(\vx_{pk} - \bar{\vx}_{p.})(\vx_{pk} - \bar{\vx}_{p.})^{T})}
\end{equation}

where $\bar{\vx}_{p.}$ is the center of each cell type cluster, and $\bar{\vx}_{..}$ is the center of all data points.

Let $x$ denote the embedded coordinate along a specific dimension. The SNR metric for that axis is therefore:

\begin{equation} \label{eqA45}
SNR = \frac{\Sigma_{p=1}^{q}n_{p}(\bar{x}_{p.} - \bar{x}_{..})^{2}}{\Sigma_{p=1}^{q}\Sigma_{k=1}^{n_{p}}(x_{pk} - \bar{x}_{p.})^{2}}
\end{equation}

\subsubsection{Explained Variance (EV)}
\label{appendixEV}

We want to know whether the variation of a specific dimension is strongly explained by that of a specific feature. Therefore, we measure, for each axis, how much of the total explained variance is explained by the variance of the feature $i$ or $j$. Formally, given the embedded coordinate $x_{ijk}$, we calculate the EV as the following:

\begin{equation} \label{eqA46}
EV_{i} = \frac{\sum_{i=1}^{a}\sum_{j=1}^{b}n_{ij} (\bar{x}_{i..} - \bar{x}_{...})^2}{\sum_{i=1}^{a} \sum_{j=1}^{b} \sum_{k=1}^{n_{ij}} (x_{ijk} - \bar{x}_{...})^2}
\end{equation}

\begin{equation} \label{eqA47}
EV_{j} = \frac{\sum_{i=1}^{a}\sum_{j=1}^{b}n_{ij} (\bar{x}_{.j.} - \bar{x}_{...})^2}{\sum_{i=1}^{a} \sum_{j=1}^{b} \sum_{k=1}^{n_{ij}} (x_{ijk} - \bar{x}_{...})^2}
\end{equation}

where $\bar{x}$ is the average of $x_{ijk}$ over the indices indicated by the dots.

\subsubsection{Mutual Information (MI)}
\label{appendixMI}

The MI between a discriminant axis $\vu$ and a feature quantifies how much information of the feature is obtained by observing data projected along that axis. It is calculated as the MI between data representations along the axis $\vy = \vu^{T}\mX$ and feature labels of the data $\vf$, where $\mX$ is the original gene expression matrix:

\begin{equation} \label{eqA48}
\begin{split}
I(\vy, \vf) &= H(\vy) + H(\vf) - H(\vy, \vf) \\
&= -\sum_{y \in Y} {p(y)\log_2 p(y)} -\sum_{f \in F} {p(f)\log_2 p(f)} -\sum_{y \in Y}\sum_{f \in F} {p(y,f)\log_2 p(y,f)}
\end{split}
\end{equation}

Here $H$ indicates entropy. To calculate $H(\vy)$ and $H(\vy, \vf)$, we discretize $\vy$ into 10 bins.

\subsubsection{Modularity}
\label{appendixModularity}

Ridgeway and Mozer (2018) argued that in a modular representation, each axis should depend on at most a single feature. Following the arguments in their paper, the modularity score is computed as follows: we first calculate the MI between each feature and each axis ($m_{if}$ denotes the MI between one axis $i$ and one feature $f$). If an axis is perfectly modular, it will have high mutual information for only one feature and zeros for the others, we therefore compute a template $t_{if}$ as the following:

\begin{equation} \label{eqA49}
    t_{if}= 
\begin{cases}
    \theta_{i},      & \text{if } f = \argmax_{g}{m_{ig}}\\
    0,               & \text{otherwise}
\end{cases}
\end{equation}

where $\theta_{i} = \max_{g}{m_{ig}}$. We then calculate the deviation from the template as:

\begin{equation} \label{eqA50}
\delta_{i} = \frac{\sum_{f}(m_{if}-t_{if})^2}{\theta_{i}^2(N-1)}
\end{equation}

where $N$ is the number of features. The modularity score for the axis $i$ is $1-\delta_{i}$. The mean of $1-\delta_{i}$ over $i$ is defined as the overall modularity score.

\subsubsection{Silhouette Score}
\label{appendixSilhouette}

The Silhouette score can be used to measure separation between clusters. The formula for the Silhouette score of one sample point is given by:

\begin{equation} \label{eqA51}
s(i) = \frac{b(i) - a(i)}{\max\{a(i), b(i)\}}
\end{equation}

where:
$a(i)$ is the average distance from the $i$th sample to the other samples in the same cluster, and $b(i)$ is the smallest average distance from the $i$th sample to the samples in the other clusters.

The Silhouette score ranges from -1 to 1. If the score is close to 1, it means that the sample is appropriately clustered. If the score is close to -1, then by the same logic, one could tell that it would be more appropriate if the data was clustered in its neighbouring cluster. The overall Silhouette score is the average silhouette score of all samples, serving as a measure of how appropriately the data have been separated and clusters.

\subsection{E. Implementation details of annotation perturbation}
\label{appendixImplementationOfAnnotationPerturbation}

We conducted an annotation perturbation analysis to assess the impact of incorrect phenotypic labeling on the performance of our FLDA method. For this, we utilized the T4/T5 neurons dataset and introduced three distinct types of perturbations to the original labels:

Firstly, we interchanged the phenotype labels of T4a neurons with one of the seven remaining types (T4b, T4c, T4d, T5a, T5b, T5c, T5d). In this case, phenotype labels for two cell types were erroneous, while the total number of cell type clusters remained the same. We maintained two levels for dendritic phenotypes (T4/T5) and four levels for axonal phenotypes (a/b/c/d). As per our methodological setup, we used one dimension to represent the dendritic feature and three dimensions for the axonal feature.

Secondly, we combined the axonal phenotypic level a with another level (b/c/d) to create an inaccurate new level (a+b/a+c/a+d). Under this scenario, there were three axonal phenotypes, we employed two dimensions to represent the axonal feature.

Lastly, we arbitrarily divided each of the four axonal lamination labels (a/b/c/d) into two distinct levels. For example, among neurons with the original axonal level a, we designated some of them with a level a1, while the rest were assigned a level a2. This resulted in eight axonal phenotypes (a1/a2/b1/b2/c1/c2/d1/d2), and following our methodology, we designated seven dimensions to account for the axonal feature.

In each scenario, we applied FLDA on the T4/T5 neurons dataset using the altered annotations and measured the respective metrics. These metrics were subsequently compared with those derived from the original annotation to assess the robustness of our method to mislabeling.

\bibliography{neurips}
\bibliographystyle{unsrt}

\section{Code Availability}
\label{code}

FLDA analysis was performed in Python, and the code and documentation are available at https://github.com/muqiao0626/FLDA-in-ComputBiol.

\section{Acknowledgement}
\label{acknowledgement}

We thank Jialong Jiang, Yuxin Chen, Oisin Mac Aodha, Matt Thomson, Lior Pachter, Andrew McMahon, Christof Koch, Yu-li Ni, Karthik Shekhar, Joshua R. Sanes, Tony Zhang, and Markus Meister for their helpful discussions and comments. 

\section{Supplementary Material}
\label{supplementary}

\begin{table}[h]
\centering
\begin{tabular}{|l|l|l|l|l|l|}
\hline
Sigma & FLDA        & 2LDAs       & LDA         & CCA         & PCA         \\ \hline
2     & 0.905050961 & 0.899968945 & 0.904654957 & 0.904642723 & 0.862205691 \\ \hline
4     & 0.809044235 & 0.799580694 & 0.80809655  & 0.808670943 & 0.70561478  \\ \hline
6     & 0.70898633  & 0.697440319 & 0.707543347 & 0.708353597 & 0.524329239 \\ \hline
8     & 0.624518364 & 0.613810816 & 0.622878691 & 0.624528348 & 0.33756613  \\ \hline
10    & 0.535243253 & 0.526429522 & 0.532720359 & 0.534676146 & 0.145181255 \\ \hline
\end{tabular}
\caption{Average Silhouette scores for FLDA and other models.}
\label{tab:table1}
\end{table}

\begin{table}[h]
\centering
\begin{tabular}{|l|l|l|l|l|l|}
\hline
Sigma & FLDA        & 2LDAs       & LDA         & CCA         & PCA         \\ \hline
2     & 0.67947216  & 1.540404677 & 0.760401511 & 1.453615189 & 0.045516276 \\ \hline
4     & 0.673864269 & 1.524357891 & 0.768786931 & 1.277295494 & 0.045400095 \\ \hline
6     & 0.670872188 & 1.523296094 & 0.763518047 & 1.225833297 & 0.046381855 \\ \hline
8     & 0.674873614 & 1.526811552 & 0.761365056 & 1.207562256 & 0.045538688 \\ \hline
10    & 0.679605055 & 1.521809649 & 0.759853601 & 1.190890384 & 0.045457077 \\ \hline
\end{tabular}
\caption{Average execution time (in seconds) for FLDA and other models.}
\label{tab:table2}
\end{table}

\end{document}